\documentclass{article}

\usepackage[final]{neurips_2021_ml4ps}

\usepackage[utf8]{inputenc} % allow utf-8 input
\usepackage[T1]{fontenc}    % use 8-bit T1 fonts
\usepackage[hyphens]{url}
\usepackage[hidelinks]{hyperref}
\usepackage{booktabs}       % professional-quality tables
\usepackage{amsfonts}       % blackboard math symbols
\usepackage{nicefrac}       % compact symbols for 1/2, etc.
\usepackage{microtype}      % microtypography
\usepackage{xcolor}         % colors
\usepackage[pdftex]{graphicx}
\usepackage{subfigure}
\title{Extending turbulence model uncertainty quantification using machine learning}

\author{%
  Marcel Matha and Christian Morsbach \\
  Institute of Propulsion Technology,  German Aerospace Center (DLR)\\
  Linder Höhe, 51147 Cologne, Germany \\
  \texttt{marcel.matha@dlr.de} and \texttt{christian.morsbach@dlr.de}\\
}

\begin{document}

\maketitle

\begin{abstract}
  In order to achieve a more virtual design and certification process of jet engines in aviation industry, the uncertainty bounds for computational fluid dynamics have to be known. 
  This work shows the application of a machine learning methodology to quantify the epistemic uncertainties of turbulence models.
  The underlying method in order to estimate the uncertainty bounds is based on an eigenspace perturbation of the Reynolds stress tensor in combination with random forests.
\end{abstract}

\section{Introduction}
  As a compromise between computational time and accuracy, Reynolds-averaged Navier-Stokes (RANS) simulation is the workhorse in the industrial design process of turbomachinery.
  The derivation of the RANS equations reveals an unclosed term, called the Reynolds stress tensor. This tensor has to be approximated in Computational-Fluid-Dynamics (CFD) simulations by applying turbulence models. 
  The prediction accuracy of the simulation is highly dependent on this kind of models, which mostly accomplished transport equations, mimicking main turbulence flow physics.
  Although RANS-based models, such as linear eddy viscosity models, are widely used for complex engineering flows, they suffer from the inability to replicate fundamental 
  turbulent processes.
  Turbulence models are one of the main limitations in striving for reliable, environmental-friendly designs, due to general simplifying assumptions during formulation 
  of closure models. 
  These simplifications are the result of data observation and physical intuition, leading to a significant degree of epistemic uncertainty.
  
  In recent years, the interest in quantifying these uncertainties, leading to more reliable simulative results, has grown. The group of Iaccarino proposed an eigenspace perturbation framework, 
  which is based on the inability of common linear eddy viscosity models to deal with Reynolds stress tensor anisotropy \cite{Emory} \cite{Iaccarino}.
  The emergence of machine learning strategies guided the path towards data driven approaches also for the turbulence modelling community \cite{Duraisamy}. 
  Heyse et al. recently enhanced the uncertainty estimation based on the eigenspace perturbation approach by adding a data-driven method \cite{Heyse}.
  
  In order to obtain more sophisticated and trustworthy simulative results, we investigate data-driven enhancements of the Reynolds stress perturbation approach with DLR's CFD solver suite TRACE.
  TRACE is being developed by the Institute of Propulsion Technology with focus on turbomachinery and offers a parallelized, multi-block CFD solver for compressible RANS equations.\footnote{TRACE User Guide, \href{http://trace-portal.de/userguide}{trace-portal.de/userguide}}

\section{Eigenspace perturbation framework}
\label{sec_eigenspaceFramework}
\subsection{Data-free approach}
\label{sec_dataFree}
\begin{figure}[htb]
\begin{center}
\mbox{
\subfigure[Relation between unperturbed and perturbed state]{\includegraphics[scale=0.35]{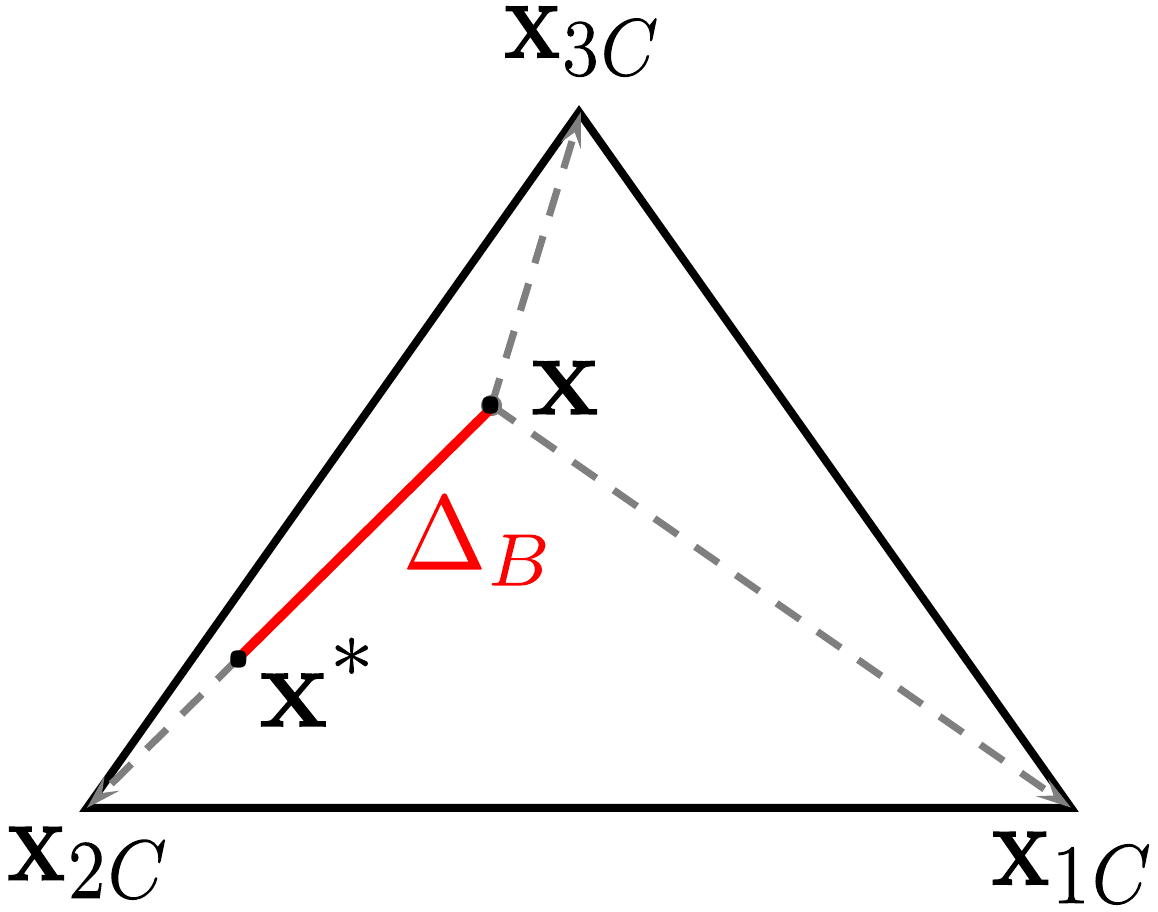}\label{baryCentric1}}}
\mbox{
\subfigure[Usage of data-driven perturbation strength]{\includegraphics[scale=0.35]{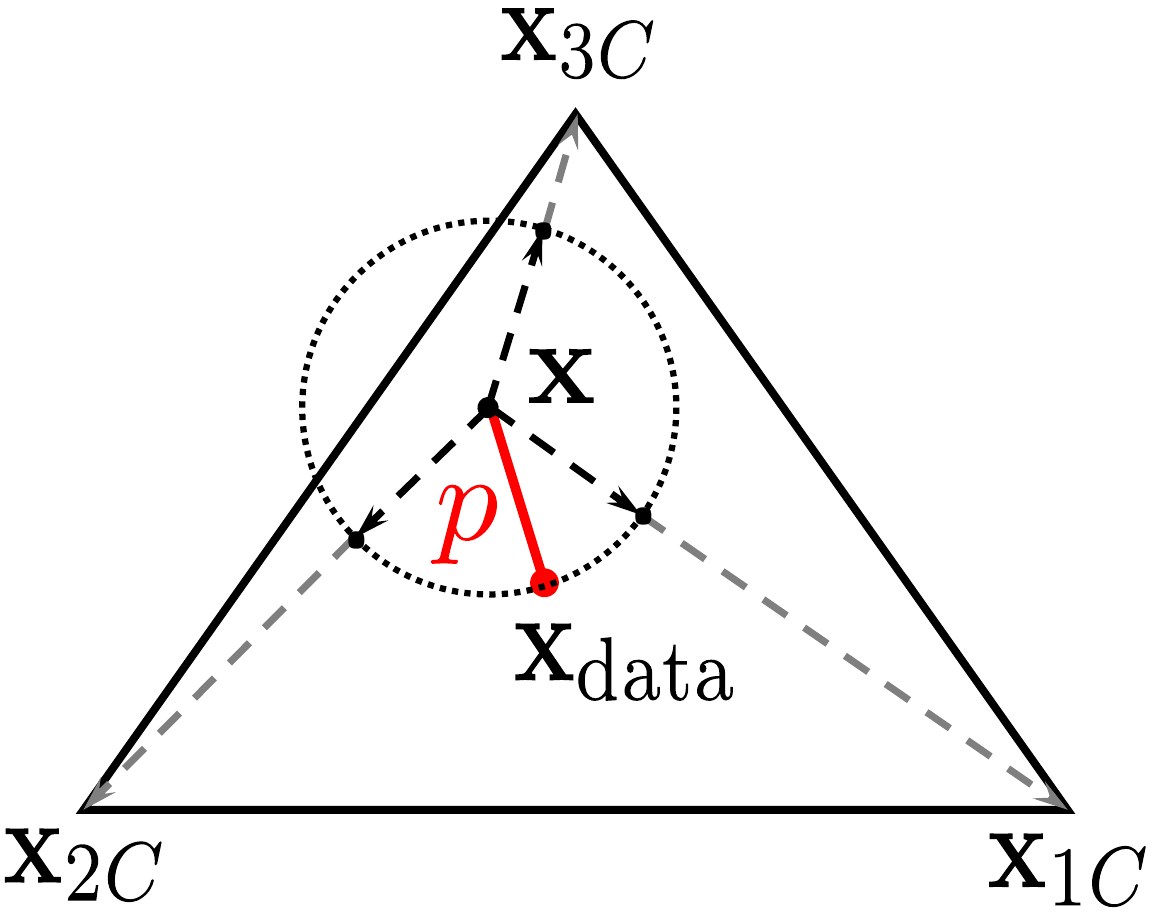}\label{baryCentric2}}}
\mbox{
\subfigure[Componentwise perturbation correction]{\includegraphics[scale=0.35]{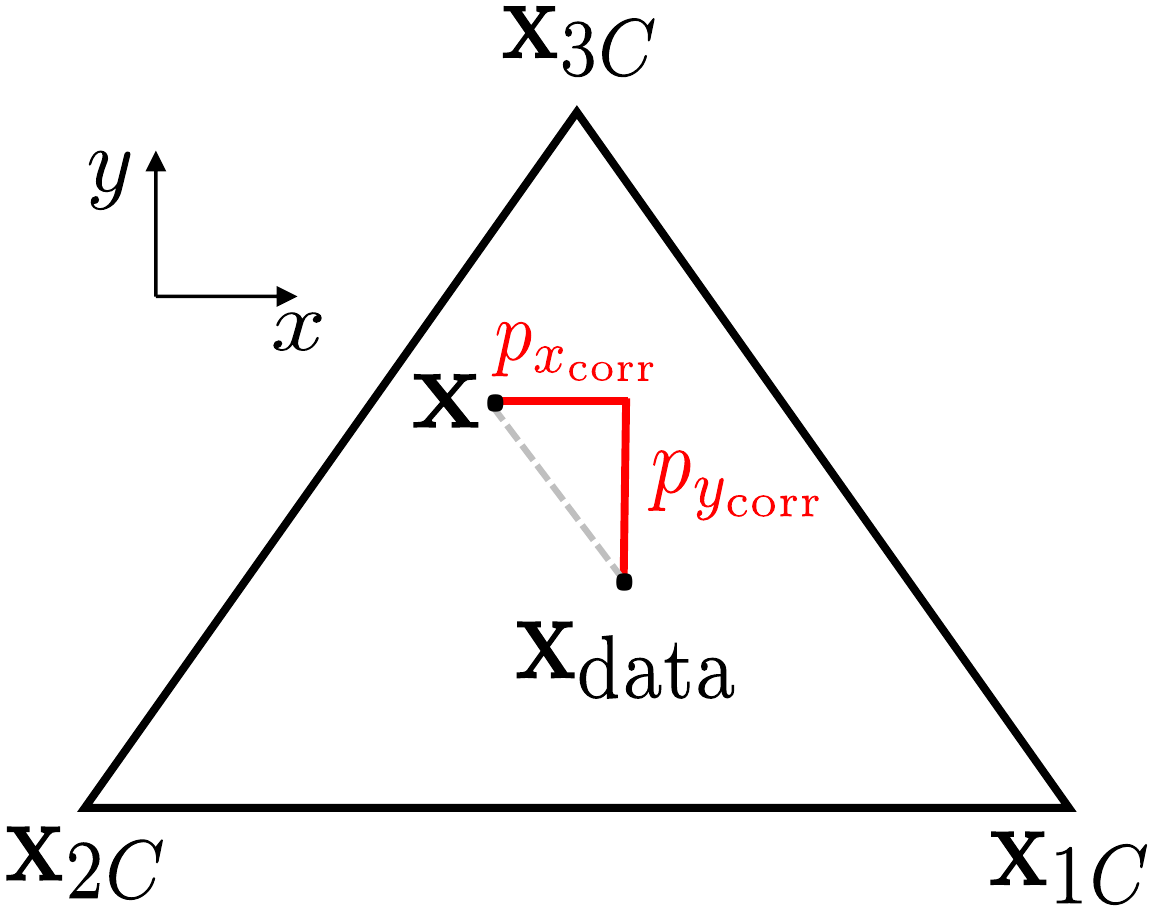}\label{baryCentric3}}}
\caption{Schematic representation of the eigenvalue perturbation approach}
\label{Bild:barycentricTriangle}
\end{center}
\end{figure}

RANS turbulence models are utilized in order to determine the Reynolds stress tensor $\tau_{ij} = \overline{u_i' u_j'}$ in terms of mean flow quantities, whereas $u'$ is the fluctuating part of the velocity ($u = \overline{u}+ u'$).
The overbar indicates the time averaging of the flow quantities.
The symmetric Reynolds stress tensor can be expressed by applying an eigenspace decomposition as
\begin{equation}
	\label{spectralDecompositionR}
		\tau_{ij} = k \left(a_{ij} + \frac{2}{3}\delta_{ij}\right) = k \left(v_{in} \Lambda_{nl} v_{jl} + \frac{2}{3}\delta_{ij}\right) \textrm{  .}
\end{equation} 
Equation \ref{spectralDecompositionR} includes the split into the anisotropy tensor $a_{ij}$ and the isotopic part of $\tau_{ij}$, while $k = \frac{1}{2} \overline{u_i' u_i'}$ being the turbulent kinetic energy.  
The eigenspace decomposition provides the eigenvector matrix $v$ and the diagonal eigenvalue matrix $\Lambda$.
Iaccarino and co-workers proposed a strategy to perturb the eigenvalues and eigenvectors in Equation \ref{spectralDecompositionR}, resulting in a perturbed state of the Reynolds stress tensor
\begin{equation}
	\label{spectralDecompositionR*}
		\tau_{ij}^* = k \left(v_{in}^* \Lambda_{nl}^* v_{jl}^* + \frac{2}{3}\delta_{ij}\right) \textrm{  .}
\end{equation}
Every physical, realizable state of the Reynolds stress tensor, meaning to be positive semidefinite, can be mapped onto barycentric coordinates
\begin{equation}
	\label{barycentricMapping}
		\mathbf{x} = \mathbf{x}_{1c}\frac{1}{2}\left(\lambda_1-\lambda_2\right)+ \mathbf{x}_{2c}\left(\lambda_2-\lambda_3\right)+ \mathbf{x}_{3c} \frac{1}{2}\left(3 \lambda_3+1\right) \quad \mathrm{with} \quad \lambda_1\geq\lambda_2 \geq\lambda_3 \textrm{  ,}
\end{equation}
which is essentially a linear transform according to $\mathbf{x} = \mathbf{B} \lambda$ within the triangle defined by the corners $\mathbf{x}_{1c}$, $\mathbf{x}_{2c}$ and $\mathbf{x}_{3c}$ (see \ref{Bild:barycentricTriangle}) \cite{Banerjee2007} .\\
The eigenvalue perturbation makes use of the fact, that the Reynolds stress tensor is limited between the one, two and three component corner of the barycentric triangle, where the limiting states reflect 
the number of non-zero eigenvalues of the tensor.  
As shown in Figure \ref{baryCentric1}, the eigenvalue perturbation is defined as a shift in barycentric coordinates towards each of the limiting states to location $\mathbf{x}^*$, according to 
\begin{equation}
	\label{perturbationMagnitude}
		\mathbf{x}^* = \mathbf{x} + \Delta_B \left(\mathbf{x}_{(t)} -\mathbf{x}\right) \textrm{  .}
\end{equation}
The relative distance $\Delta_B$ controls the magnitude of eigenvalue perturbation towards the corner state $\mathbf{x}_{(t)} \in {\mathbf{x}_{\mathrm{1C}}, \mathbf{x}_{\mathrm{2C}}, \mathbf{x}_{\mathrm{3C}}}\}$.
The perturbed eigenvalues $\lambda_{i}^*$ can be remapped afterwards.

%\begin{equation}
%	\label{perturbedEigenvalues}
		%\lambda_{i}^* = B^{-1} \mathbf{x}^* \\ \text{,}
%\end{equation} 

The creation of the perturbed eigenvector matrix $\Lambda^*$ is purely motivated on the manipulation of the turbulent production term $P_k=-\tau_{ij}\frac{\partial u_i}{\partial x_j}$ and is not discussed further during this extended abstract.
%Changing the alignment of the eigenvectors of the Reynolds stress tensor and the strain rate tensor $S_{ij}$ limits the production term to a maximum or minimum value.
Taking only the eigenvalue perturbation into account, a designer needs to run three distinct RANS simulations in order to get the full estimate of the turbulence model uncertainty bounds, 
since the methodology seeks to sample from the three extremal states in the barycentric map.
\subsection{Data-driven approach}
\label{sec_dataDriven}
The data-free approach is a purely physics-based methodology, aiming for understandable uncertainty bounds for the turbulence modelling community. One of the major drawbacks of the proposed method is the fact, that a user 
has to choose the degree of uncertainty by selecting $\Delta_B$ before each investigation. 
The perturbation amplitude, closely related to the degree of uncertainty, has only to become significant in flow regime, which contravene the assumptions during formulation of the 
turbulence model. Thus, enabling a spatially varying perturbation of the Reynolds stress tensor seems to be worthwhile.\\
The prediction of the correct anisotropy was investigated in an a priori study using random forests by Wang et al. \cite{Wang}. 
Heyse et al. propose a strategy to combine a random forest model with the forward eigenvalue perturbation approach \cite{Heyse}. Physical flow features are extracted to train a random forest 
in order to predict the local perturbation strength $p = |\mathbf{x}_{\mathrm{Data}} - \mathbf{x}_{\mathrm{RANS}}| = |\mathbf{x}^*_{\mathrm{RANS}} - \mathbf{x}_{\mathrm{RANS}}|$ (see Figure \ref{baryCentric2}). 
Forward propagating CFD simulations follow training the model, where the predicted perturbation strength is used to modify the Reynolds stress towards the same three limiting states as in the data-free approach. \\
For the sake of interpretability and usability, an ensemble learning random regression forest is chosen in the presented work as well \cite{Breiman}.
The python library \textit{scikit-learn} is used to train the random forest and evaluate its prediction.
We will investigate the proposed combination of a machine learning model and the perturbation approach within TRACE.
The aim is to go one step further and merge the work of Wang et al. \cite{Wang} and Heyse et al. \cite{Heyse}.\\
As a first step a machine learning model should predict the componentwise perturbation strength $\mathbf{p}_{\mathrm{corr}} = (p_{x_{\mathrm{corr}}}$, $p_{y_{\mathrm{corr}}})$ with respect to data, as illustrated in Figure \ref{baryCentric3}.
The effect of applying corrected eigenvalues of the Reynolds stresses can be observed by propagating them through the solver. \\
Furthermore, the orientation of the Reynolds stress tensor in terms of mutually orthogonal eigenvectors, can be described as a rigid body rotation with three degrees of freedom.
RANS results show a discrepancy to the orientation of the eigenvectors obtained from high-fidelity data, e.g. scale resolving simulations.
Consequently, intrinsic Tait-Bryan angles $\alpha, \beta, \gamma$ in the $z-y'-x''$ convention according to Rothmayr and Hodges \cite{Roithmayr2016DynamicsTA} will be used to describe the eigenvector rotation
from RANS results towards the data.  
The combination of corrected eigenvalues and eigenvectors completes a predictive framework to model anisotropy discrepancy and propagate the predicted Reynolds stresses to mean flow quantities.
In terms of selected flow features, we extended physical quantities to the proposed exhaustive invariant feature list in the work of Ling et al. \cite{Ling} and Wang et al. \cite{Wang}.
\section{Channel flow results}
\label{sec_results}
The fully developed turbulent channel flow offers a possibility to demonstrate the coupling of the uncertainty quantification (UQ) perturbation approach with a machine learning model. 
The discrepancies between RANS and direct numerical simulations (DNS) of Lee and Moser \cite{lee_moser_2015} at different Reynolds numbers $Re_\tau$ serve as training and testing data for the random forests.\\
\begin{table}[htb]
  \caption{Selected hyperparameters for the random forest regressors}
  \label{hyperparameters}
  \centering
  \begin{tabular}{llll}
    \toprule
                       &         \multicolumn{3}{c}{Target quantities}                   \\
    \cmidrule(r){2-4}
    Hyperparameter     & $p$     & $\mathbf{p}_{\mathrm{corr}}$ & $\mathbf{p}_{\mathrm{corr}}$ + $\alpha, \beta, \gamma$\\
    \midrule
    max tree depth       & 6  & 9  & 9 \\
    min sample count     & 6  & 4  & 4 \\
    max active features  & 3  & 3  & 3 \\
    number of trees      & 30 & 15 & 30\\
    \bottomrule
  \end{tabular}
\end{table}

Before the actual training of the model was conducted, the impact of four different hyperparameters on the accuracy of the random regression forest model is evaluated: 
the maximum tree depth, the minimum sample count, the maximum number of active features and the number of trees. The selected parameters for the respective target quantities are
based on hyperparameter studies (see Table \ref{hyperparameters}).
The random forest regression models are trained on $Re_\tau \in \{180, 550, 2000, 5200\}$ using the mean squared error to determine the quality of each split, 
whereas the models are evaluated in forward CFD-simulations at $Re_\tau = 1000$. 
The discrepancies with respect to barycentric coordinates are moderate in the channel center and start to increase close to the wall due to strong anisotropy of turbulence.
The two-equation, linear eddy viscosity Menter SST $k$-$\omega$ turbulence model is considered as the baseline model in the current study \cite{Menter}.
\begin{figure}[t]
\begin{center}
\mbox{
\subfigure[Streamwise velocity profile]{\includegraphics[scale=0.45, trim={0 0 1.5cm 1.3cm},clip]{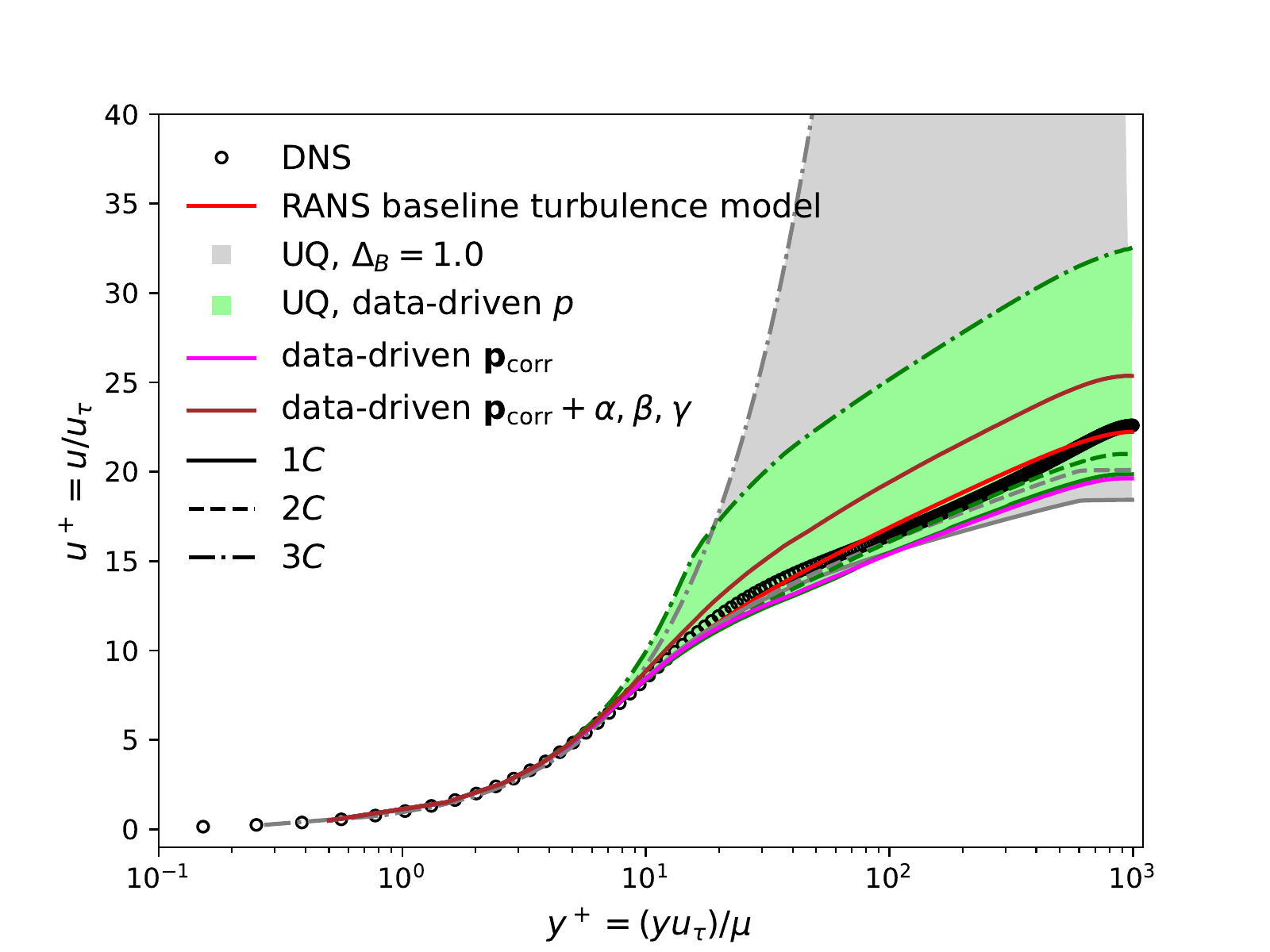}\label{velocity_UQ}}}
\mbox{
\subfigure[Barycentric coordinates]{\includegraphics[scale=0.45, trim={1.8cm 0 1.8cm 1.65cm},clip]{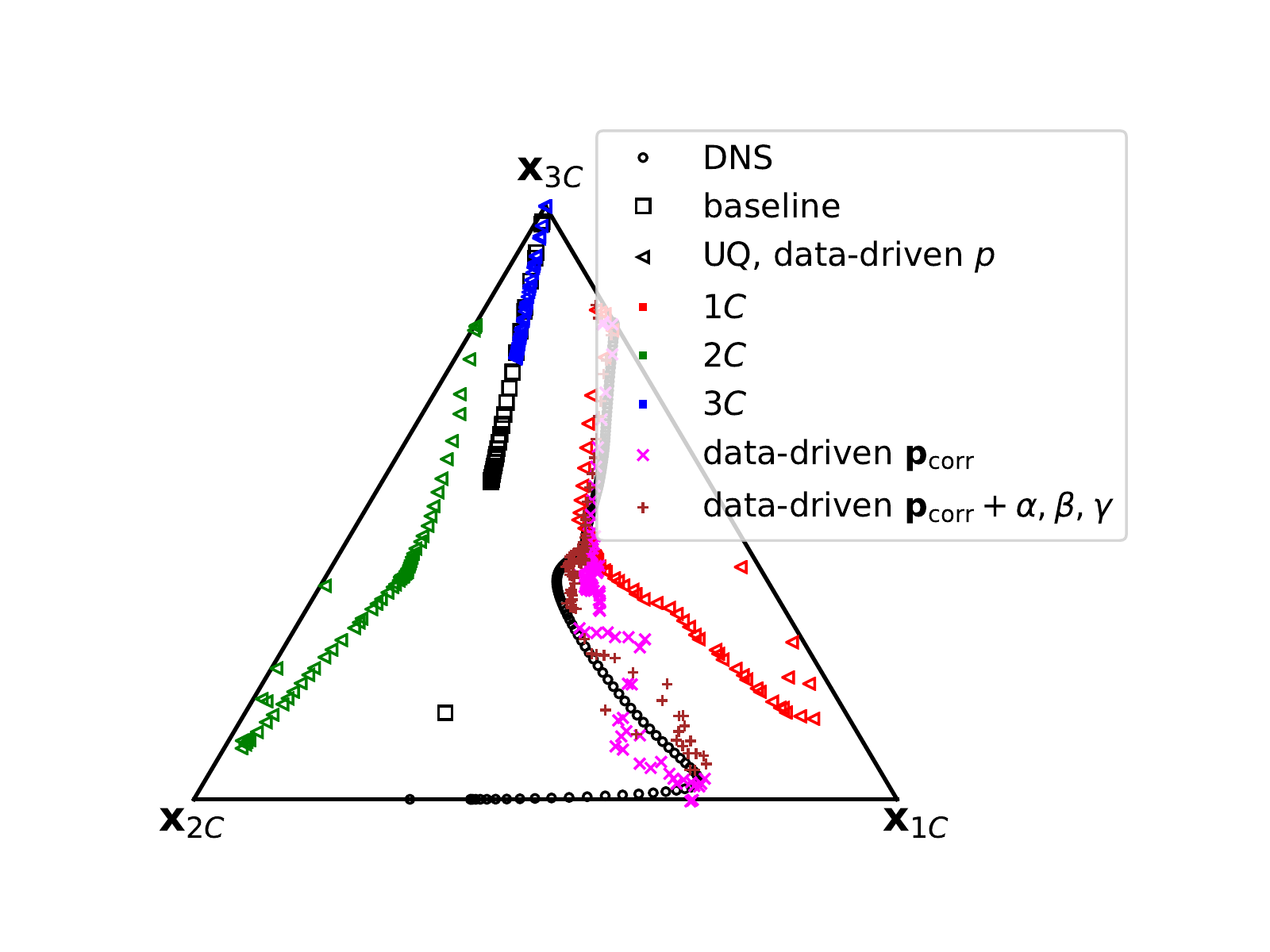}\label{baryTriangle_UQ}}}
\caption{Performance of random forest models in combination with the perturbation uncertainty quantification approach at channel flow simulations of $Re_\tau = 1000$}
\label{Bild:modelOutput_1}
\end{center}
\end{figure}
The advantage of the data-driven UQ approach, determining a relative perturbation strength $p$ can be seen in Figure \ref{Bild:modelOutput_1}. The predicted uncertainty bounds in Figure \ref{velocity_UQ} 
are significantly less conservative compared to the data-free method with a chosen $\Delta_B = 1.0$, while the traces of the barycentric coordinates look plausible for each of the perturbed solutions 
(see Figure \ref{baryTriangle_UQ}).
As expected, the baseline model aligns itself with the plane strain line, whereas the perturbed solutions tend to the limiting states of turbulence in the corners.
Additionally, the random forest model, predicting the perturbation direction vector $\mathbf{p}_{\mathrm{corr}}$, was evaluated in forward CFD solver mode. Although it struggles to 
predict the $x-$component of the perturbation, the resulting eigenvalues of the Reynolds stress tensor in the simulation run are satisfying, as illustrated in Figure \ref{baryTriangle_UQ}. 
Unfortunately, guaranteeing eigenvalues close to the truth does not seem to be sufficient in order to obtain the correct quantity of interest, in this case the velocity profile.
\begin{figure}[h]
\begin{center}
\mbox{
\subfigure[rotation around $z$-axis ($\alpha$)]{\includegraphics[scale=0.29, trim={0.1cm 0.5cm 0.39cm 0.35cm},clip]{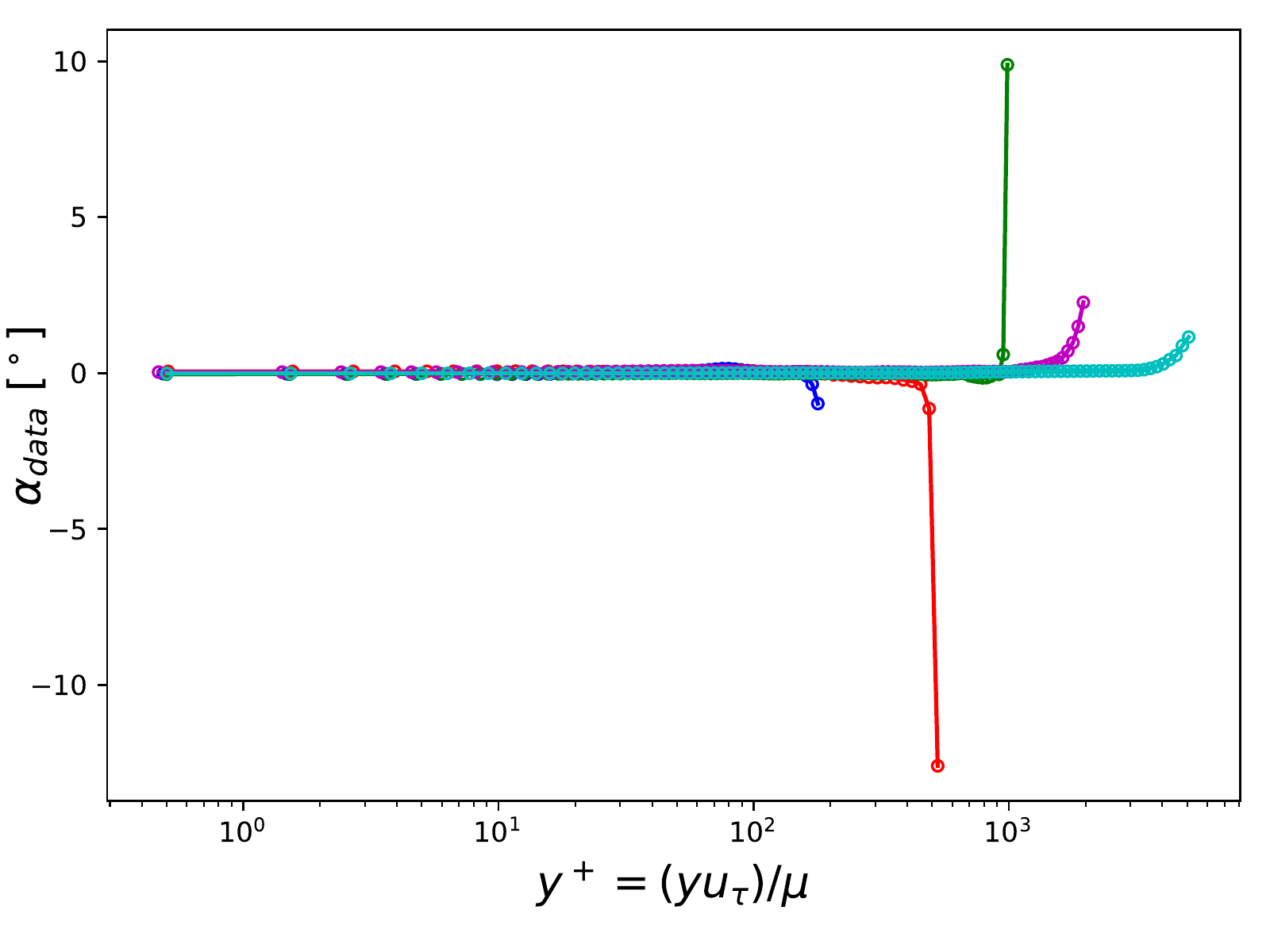}\label{eulerAlpha}}}
\mbox{
\subfigure[rotation around $y'$-axis ($\beta$)]{\includegraphics[scale=0.29, trim={0.42cm 0.5cm 0.39cm 0.35cm},clip]{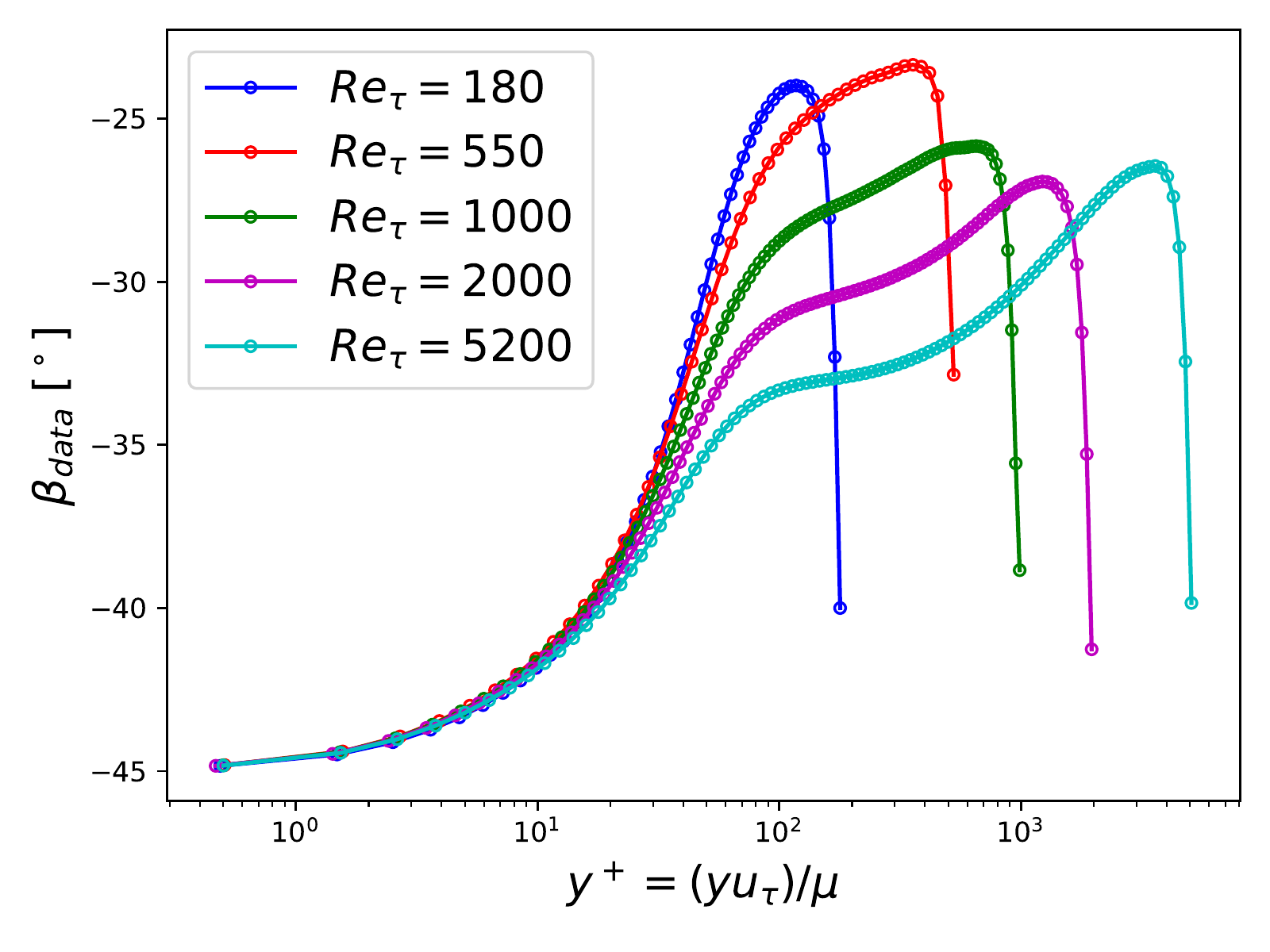}\label{eulerBeta}}}
\mbox{
\subfigure[rotation around $x''$-axis ($\gamma$)]{\includegraphics[scale=0.29, trim={0.42cm 0.5cm 0.39cm 0.35cm},clip]{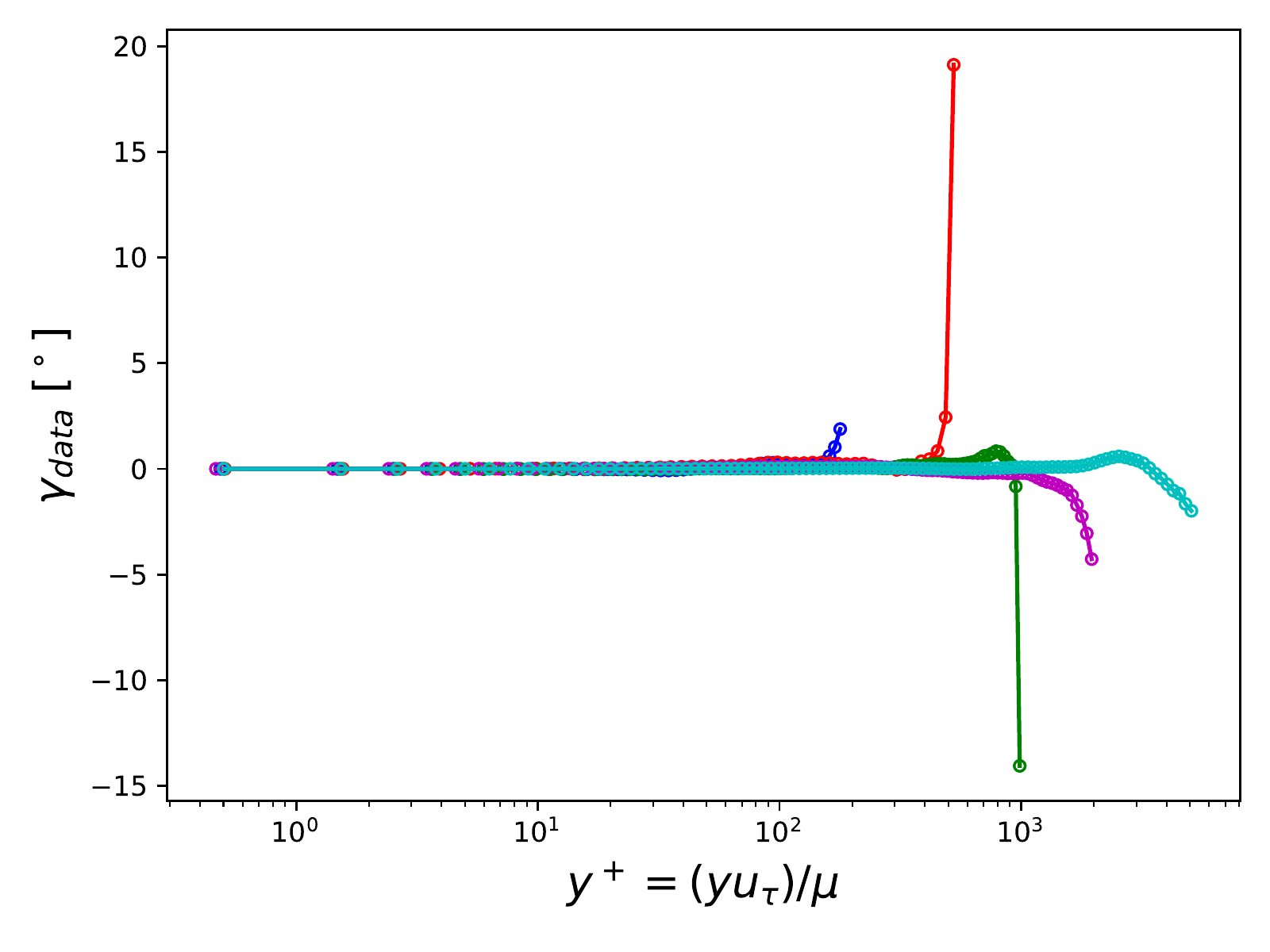}\label{eulerGamma}}}
\mbox{
\subfigure[model accuracy for $\alpha$]{\includegraphics[scale=0.31, trim={0.3cm 0 1.55cm 1.45cm},clip]{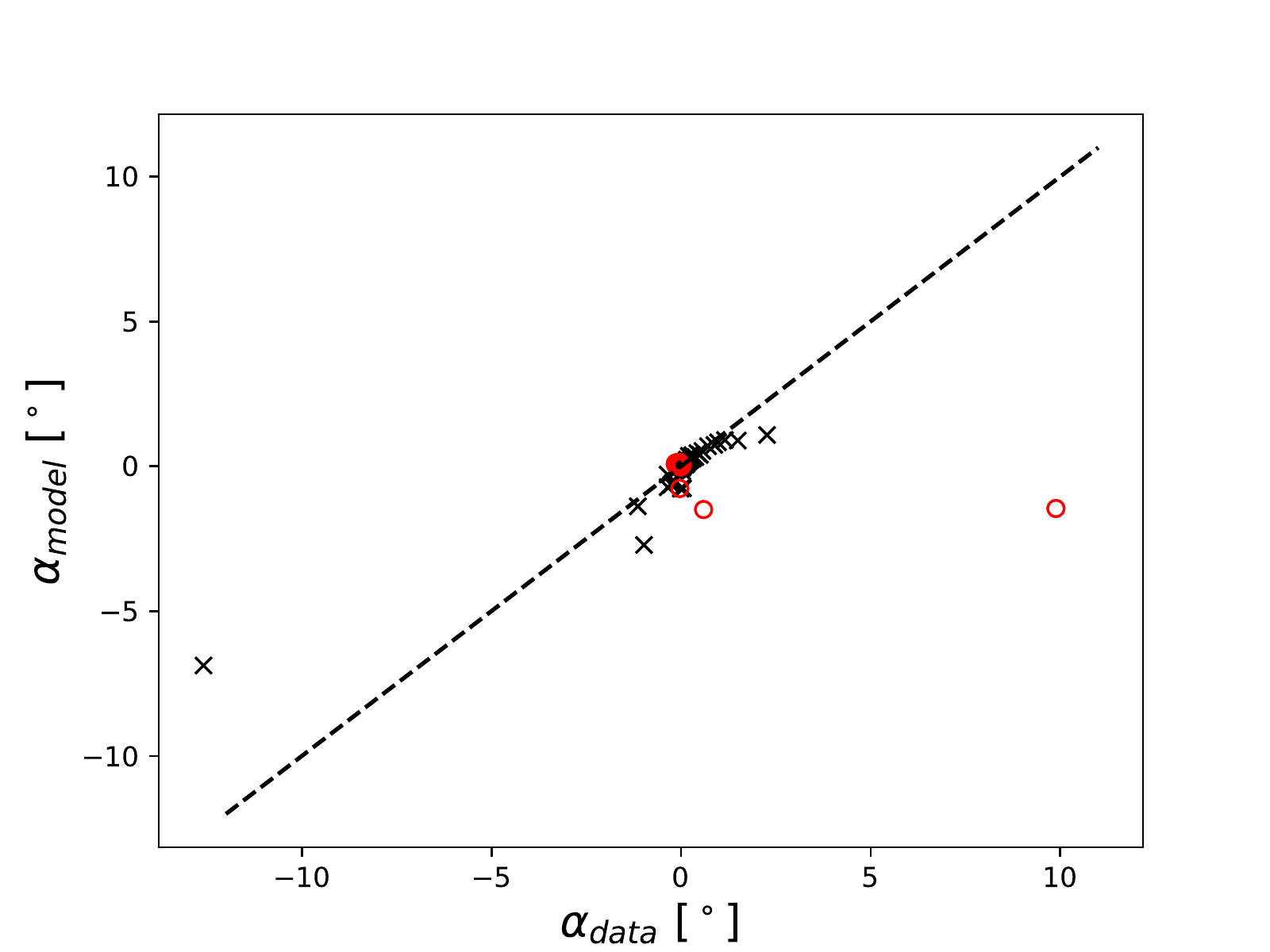}\label{eulerAlphaTrain}}}
\mbox{
\subfigure[model accuracy for $\beta$]{\includegraphics[scale=0.31, trim={0.18cm 0 1.55cm 1.45cm},clip]{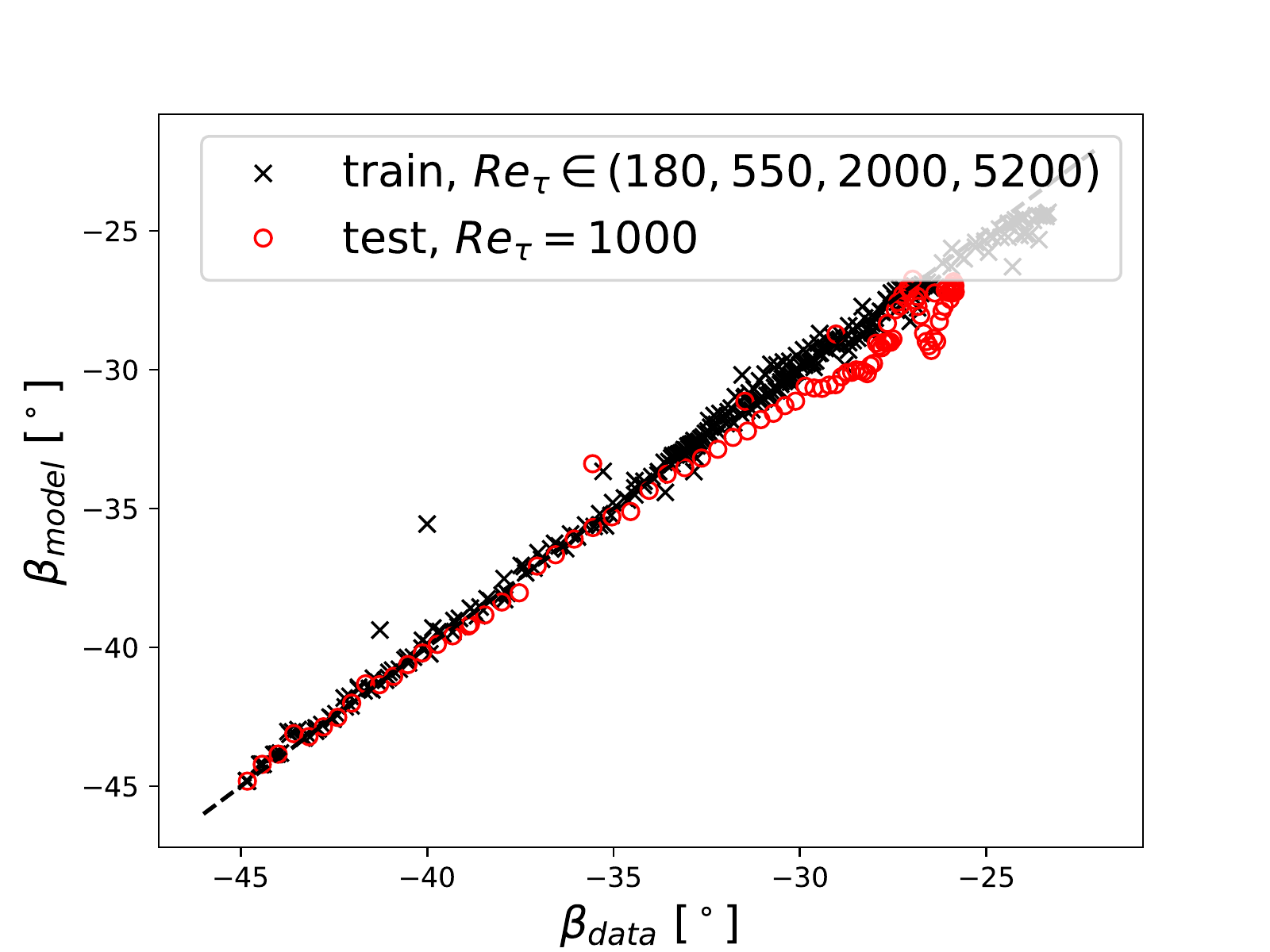}\label{eulerBetaTrain}}}
\mbox{
\subfigure[model accuracy for $\gamma$]{\includegraphics[scale=0.31, trim={0.3cm 0 1.5cm 1.45cm},clip]{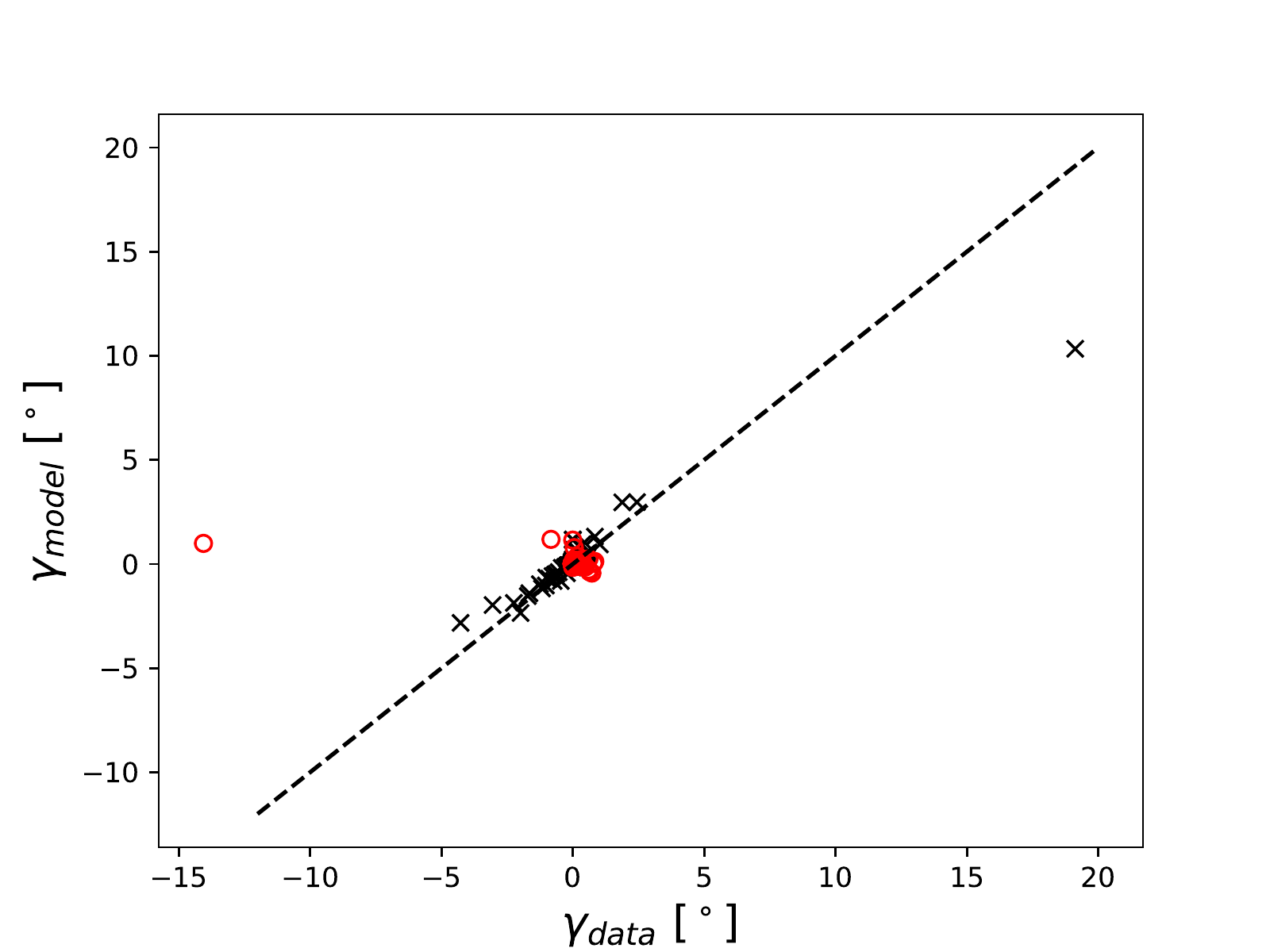}\label{eulerGammaTrain}}}
\caption{Overview of the eigenvector rotation (RANS $\rightarrow$ DNS) for channel flow}
\label{Bild:modelOutput_2}
\end{center}
\end{figure}

\begin{figure}[htb]
\begin{center}
\mbox{
\subfigure[Main components of the Reynolds stress tensor]{\includegraphics[scale=0.31, trim={0 0 1.6cm 1.3cm},clip]{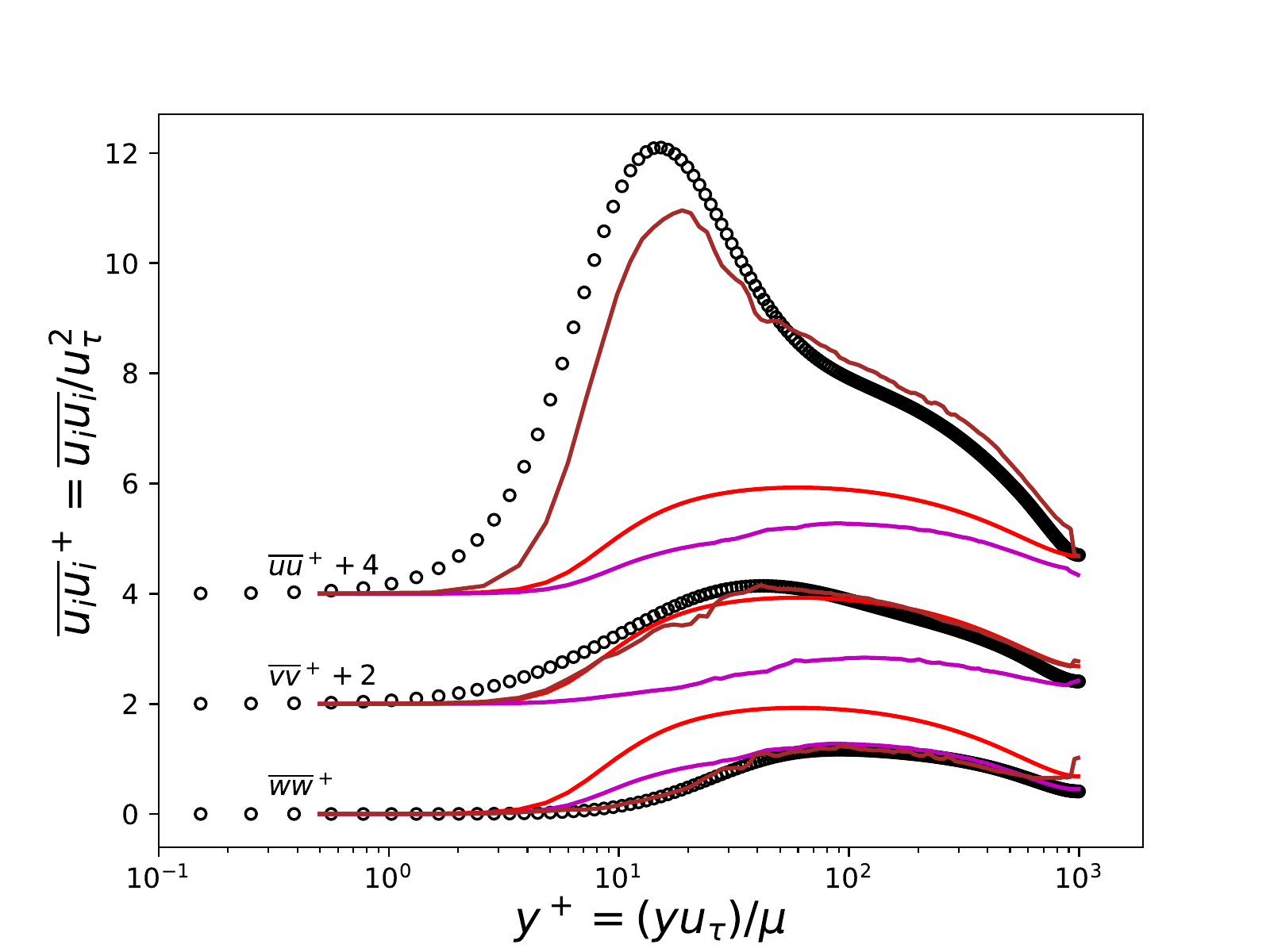}\label{uiujPerturbed}}}
\mbox{
\subfigure[Shear stress component of the Reynolds stress tensor]{\includegraphics[scale=0.31, trim={0 0 1.6cm 1.3cm},clip]{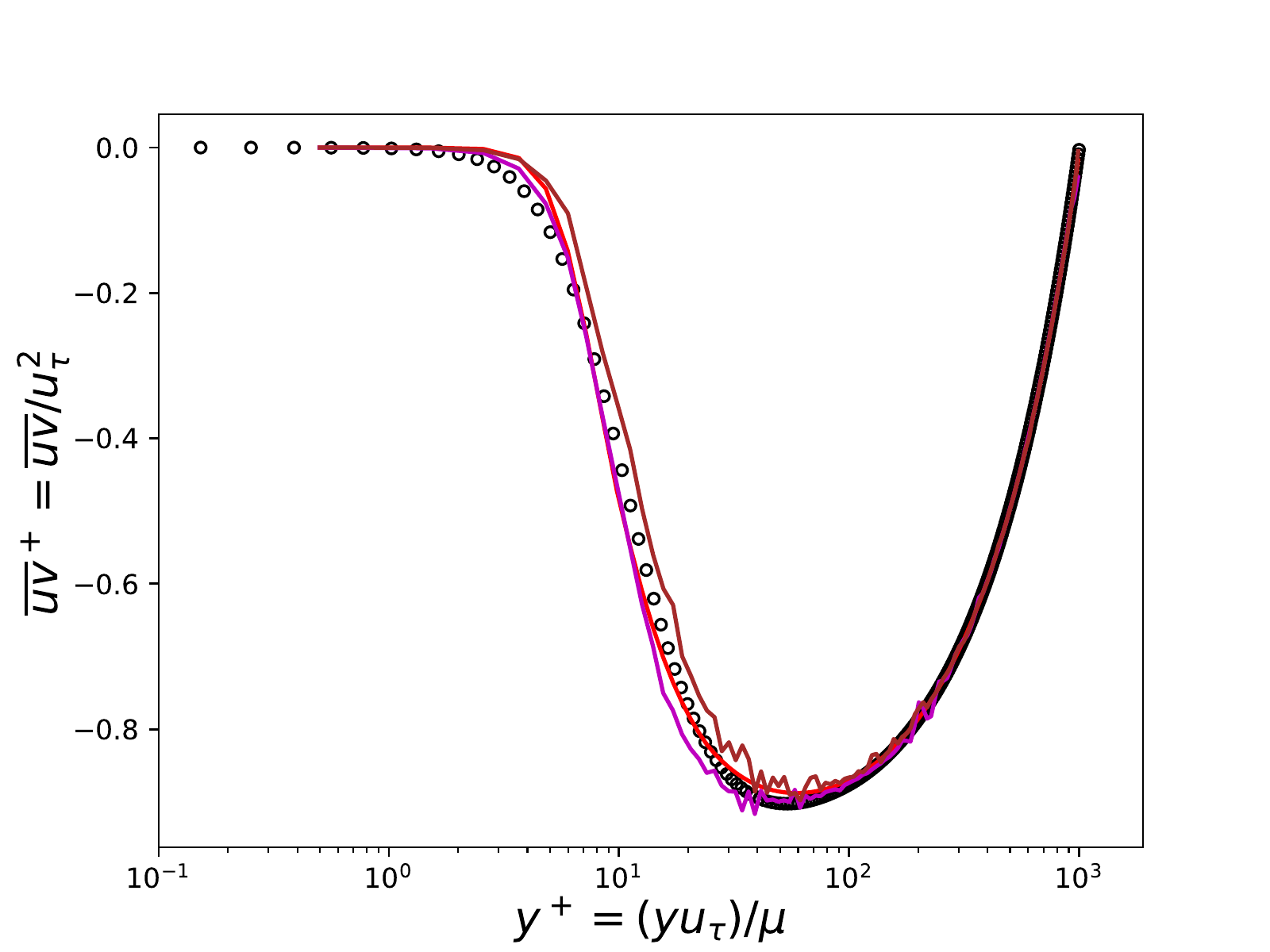}\label{uvPerturbed}}}
\mbox{
\subfigure[Turbulent kinetic energy]{\includegraphics[scale=0.31, trim={0.7cm 0 1.5cm 1.3cm},clip]{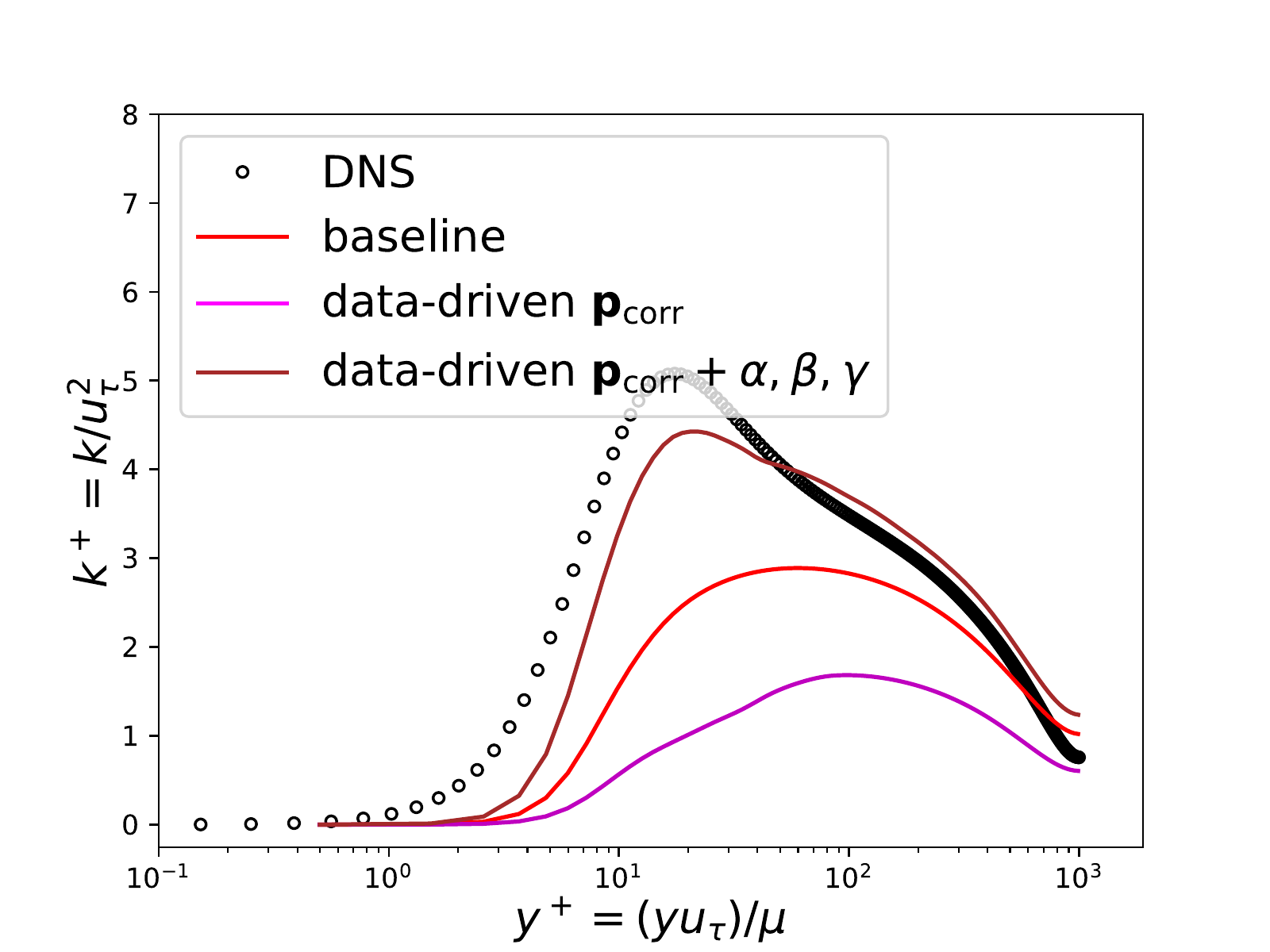}\label{tkeChannel}}}
\caption{Evaluation of model predicted turbulent quantities at channel flow of $Re_\tau = 1000$}
\label{Bild:modelOutput_1}
\end{center}
\end{figure}

Moreover, we start to explore the discrepancy field regarding the eigenvector orientation of the RANS and the DNS data in Figure \ref{eulerAlpha}, \ref{eulerBeta} and \ref{eulerGamma}. 
Unlike the differences in barycentric coordinates, the eigenvector misalignment seems to be significant in the log-law region.
A random forest regressor was trained to predict the Tait-Bryan angles, as discussed above. 
Except for some outliers, the model is able to predict the correct rotation angles within an acceptable range (see Figure \ref{eulerAlphaTrain}, \ref{eulerBetaTrain} and \ref{eulerGammaTrain}). 
A combination of propagating true eigenvalues and eigenvectors ($\mathbf{p}_{\mathrm{corr}} + \alpha, \beta, \gamma$) in the forward CFD application led to an improved prediction of the Reynolds stresses, as illustrated in Figure \ref{uiujPerturbed}, 
although actually only the anisotropy tensor was corrected without modifying the turbulent kinetic energy explicitly (see Equation \ref{spectralDecompositionR*}).
Due to the continual response of the equations to the changed Reynolds stresses, the turbulence model produces more accurate turbulent kinetic energy compared with the DNS data (see Figure \ref{tkeChannel}).

It might be possible, that the non-smoothness of the model prediction causes an issue on the derivatives of the Reynolds stresses, entering the RANS equations, leading to a descrepancy of the flow field \cite{Wang}.  
In order to answer the research question, whether a propagation of corrected Reynolds stresses is able to improve the accuracy of quantities of interests, we propagated interpolated Reynolds stresses of the DNS through the
RANS equations and observed a perfect match for the velocity profile at $Re_\tau = 1000$. Additionally, adding random noise to these interpolated Reynolds stresses and not noticing significant change in the velocity profile (not shown here),
refute the non-smoothness theory. 
Therefore, the different behavior of the shear stress component between $10 \leq y^+ \leq 40$ in Figure \ref{uvPerturbed} is accountable for the misalignment of the velocity profile for the simulations containing 
a machine learned correction of the eigenvalues ($\mathbf{p}_{\mathrm{corr}}$) and of the entire anisotropy tensor ($\mathbf{p}_{\mathrm{corr}} + \alpha, \beta, \gamma$).
Further work needs to deal with an additional correction for the turbulent kinetic energy based on a machine learning model, which potentially lead to more accurate quantities of interest.

\section{Conclusion}
\label{sec_concluison}
This work presents the possibility to estimate data-driven uncertainty bounds for turbulence model with state-of-the-art methods in DLR's CFD solver TRACE. 
Moreover, recent approaches in literature acted as thought-provoking impulse to train machine learning models on parameters of the initial perturbation approach.  
The existing concerns were confirmed, that propagating the Reynolds stress corrections is challenging and might not easily lead to correct mean flow quantities of interest.
Nevertheless, getting to know the turbulence model uncertainties and especially localize flow regions, where those are present, will help making design exploration and certification more reliable.

\paragraph{Impact statement} 
Complex engineering designs always feature turbulent flows. Although computational power is increasing massively during the past years, scale resolving simulations for design optimization studies 
seems to be still a long way off.
RANS simulation, whose accuracy is heavily dependent on turbulence models, will remain state-of the art in the upcoming years.
%Thus, on the way towards a virtual certification process - not only in aircraft and turbomachinery industry - uncertainty quantification of CFD simulations is the tool to be. 
The present work aims to consolidate the arisen methods in the field of turbulence model uncertainty quantification in combination with machine learning methods. 
Additionally, designing and implementing a framework to easily conduct uncertainty estimation for turbulence models was a major goal of this work.  
The authors do not see any ethical concern regarding the present work.

%\begin{figure}
%  \centering
%  \fbox{\rule[-.5cm]{0cm}{4cm} \rule[-.5cm]{4cm}{0cm}}
%  \caption{Sample figure caption.}
%\end{figure}

%\begin{table}
%  \caption{Sample table title}
%  \label{sample-table}
%  \centering
%  \begin{tabular}{lll}
%    \toprule
%    \multicolumn{2}{c}{Part}                   \\
%    \cmidrule(r){1-2}
%    Name     & Description     & Size ($\mu$m) \\
%    \midrule
%    Dendrite & Input terminal  & $\sim$100     \\
%    Axon     & Output terminal & $\sim$10      \\
%    Soma     & Cell body       & up to $10^6$  \\
%    \bottomrule
%  \end{tabular}
%\end{table}

\bibliography{literatur}

\begin{thebibliography}{11}
\providecommand{\natexlab}[1]{#1}
\providecommand{\url}[1]{\texttt{#1}}
\expandafter\ifx\csname urlstyle\endcsname\relax
  \providecommand{\doi}[1]{doi: #1}\else
  \providecommand{\doi}{doi: \begingroup \urlstyle{rm}\Url}\fi

\bibitem[Banerjee et~al.(2007)Banerjee, Krahl, Durst, and Zenger]{Banerjee2007}
S.~Banerjee, R.~Krahl, F.~Durst, and C.~Zenger.
\newblock Presentation of anisotropy properties of turbulence, invariants
  versus eigenvalue approaches.
\newblock \emph{Journal of Turbulence}, 8:\penalty0 N32, 2007.
\newblock \doi{10.1080/14685240701506896}.
\newblock URL \url{https://doi.org/10.1080/14685240701506896}.

\bibitem[Breiman(2001)]{Breiman}
L.~Breiman.
\newblock Random forests.
\newblock \emph{Machine Learning}, 45:\penalty0 5--32, 10 2001.
\newblock \doi{10.1023/A:1010950718922}.

\bibitem[Duraisamy et~al.(2019)Duraisamy, Iaccarino, and Xiao]{Duraisamy}
K.~Duraisamy, G.~Iaccarino, and H.~Xiao.
\newblock Turbulence modeling in the age of data.
\newblock \emph{Annual Review of Fluid Mechanics}, 51\penalty0 (1):\penalty0
  357--377, Jan 2019.
\newblock ISSN 1545-4479.
\newblock \doi{10.1146/annurev-fluid-010518-040547}.
\newblock URL \url{http://dx.doi.org/10.1146/annurev-fluid-010518-040547}.

\bibitem[Emory et~al.(2013)Emory, Larsson, and Iaccarino]{Emory}
M.~Emory, J.~Larsson, and G.~Iaccarino.
\newblock Modeling of structural uncertainties in reynolds-averaged
  navier-stokes closures.
\newblock \emph{Physics of Fluids}, 25\penalty0 (11):\penalty0 110822, 2013.
\newblock \doi{10.1063/1.4824659}.
\newblock URL \url{https://doi.org/10.1063/1.4824659}.

\bibitem[Heyse et~al.(2021)Heyse, Mishra, and Iaccarino]{Heyse}
J.~Heyse, A.~Mishra, and G.~Iaccarino.
\newblock Estimating rans model uncertainty using machine learning.
\newblock \emph{Journal of the Global Power and Propulsion Society}, pages
  1--14, 05 2021.
\newblock \doi{10.33737/jgpps/134643}.

\bibitem[Iaccarino et~al.(2017)Iaccarino, Mishra, and Ghili]{Iaccarino}
G.~Iaccarino, A.~Mishra, and S.~Ghili.
\newblock Eigenspace perturbations for uncertainty estimation of single-point
  turbulence closures.
\newblock \emph{Physical Review Fluids}, 2, 02 2017.
\newblock \doi{10.1103/PhysRevFluids.2.024605}.

\bibitem[Lee and Moser(2015)]{lee_moser_2015}
M.~Lee and R.~D. Moser.
\newblock Direct numerical simulation of turbulent channel flow up to
  $\mathit{Re}_{{\it\tau}}\approx 5200$.
\newblock \emph{Journal of Fluid Mechanics}, 774:\penalty0 395--415, 2015.
\newblock \doi{10.1017/jfm.2015.268}.

\bibitem[Ling et~al.(2016)Ling, Jones, and Templeton]{Ling}
J.~Ling, R.~E. Jones, and J.~Templeton.
\newblock Machine learning strategies for systems with invariance properties.
\newblock \emph{J. Comput. Phys.}, 318:\penalty0 22--35, 2016.

\bibitem[Menter et~al.(2003)Menter, Kuntz, and Langtry]{Menter}
F.~Menter, M.~Kuntz, and R.~Langtry.
\newblock Ten years of industrial experience with the sst turbulence model.
\newblock \emph{Heat and Mass Transfer}, 4, 01 2003.

\bibitem[Roithmayr and Hodges(2016)]{Roithmayr2016DynamicsTA}
C.~Roithmayr and D.~Hodges.
\newblock Dynamics: Theory and application of kane's method.
\newblock 2016.

\bibitem[Wang et~al.(2018)Wang, Wu, Xiao, and Paterson]{Wang}
J.-X. Wang, J.-L. Wu, H.~Xiao, and E.~Paterson.
\newblock Physics-informed machine learning approach for augmenting turbulence
  models: A comprehensive framework.
\newblock \emph{Physical Review Fluids}, 3\penalty0 (7), Jul 2018.
\newblock ISSN 2469-990X.
\newblock \doi{10.1103/physrevfluids.3.074602}.
\newblock URL \url{http://dx.doi.org/10.1103/PhysRevFluids.3.074602}.

\end{thebibliography}

%%%%%%%%%%%%%%%%%%%%%%%%%%%%%%%%%%%%%%%%%%%%%%%%%%%%%%%%%%%%

%%%%%%%%%%%%%%%%%%%%%%%%%%%%%%%%%%%%%%%%%%%%%%%%%%%%%%%%%%%%

\appendix

%\section{Appendix}

%Optionally include extra information (complete proofs, additional experiments and plots) in the appendix.
%This section will often be part of the supplemental material.

\end{document}